\documentclass[12pt,amsfonts]{article}
\begin{document}
\def\p {{\partial}}
\def\n {{\nu}}
\def\m {{\mu}}
\def\a {{\alpha}}
\def\bt {{\beta}}
\def\f {{\phi}}
\def\th {{\theta}}
\def\g {{\gamma}}
\def\eps {{\epsilon}}
\def\e {{\psi}}
\def\la {{\lambda}}
\def\na {{\nabla}}
\def\k {\chi}
\def\bn {\begin{eqnarray}}
\def\en {\end{eqnarray}}
\title{Solutions of a particle with fractional $\delta$-potential in a fractional dimensional space}
\maketitle
\begin{center}
\textbf{Sami I. Muslih\footnote{On leave of absence from Al-Azhar
University-Gaza, Email:
smuslih@ictp.it} }\\
 $^{1}$Department~ of~  Mechanical~ Engineering,\\
 Southern Illinois University, Carbondale, Illinois - 62901, USA
\\
 \vspace{ 0.5cm}
\end{center}
\begin{abstract}
A Fourier transformation in a fractional dimensional space of order $\la$~($0<\la\leq 1$) is defined to  solve the Schr\"{o}dinger equation with
Riesz fractional derivatives of order $\a$.  This new method is applied for a particle in a fractional $\delta$-potential well defined by $V(x) =- \gamma\delta^{\la}(x)$, ~where $\gamma>0$~ and $\delta^{\la}(x)$ is the fractional Dirac delta function.
A complete solutions for the energy values and the wave functions are obtained in terms of the Fox H-functions. It is demonstrated that the eigen solutions are exist if $0< \la<\a$.
The results for $\la= 1$ and $\a=2$ are in exact agreement with those presented in the standard quantum mechanics.
\end{abstract}

\section{Introduction}
In 1918 the Mathematician Felix Hausdorff introduced the notion of fractional dimension. This concept became very important especially after
the revolutionary discovery of fractal geometry by Mandelbrot [1], where he used the concept of fractionality and worked
out the relations between fractional dimension and integer
dimension by using the scale method i.,e. $d^{\la}x=
\frac{\pi^{\la/2}|x|^{\la -1}}{\Gamma(\la/2)}dx,$ $0<\la\leq 1$. And numerous efforts has been made by researchers in various branches of science and technology [2-18]. Besides, there are other approaches to describe fractional dimension. These include, fractional calculus (a generalization of differentiation and integration to non integer order) [19] and the analytic continuation of the dimension in Gaussian integral [12, 20-22]. The later is often used in quantum field theory [21,22], and introduced in the dimensional regularization
method , $\displaystyle\int
f(x)d^{n}x=\frac{2\pi^{(n)/2}}{\Gamma(\frac{n}{2})}
\displaystyle\int_{0}^{\infty}f(x)x^{n-1}dx$(a method of removing
the divergent term in the evaluation of Feynman diagram term in
order to avoid the loop divergence).

Historically, the first example of fractional physical objects
was the Brownian  motion [1]. In quantum physics the first successful attempt
applying of fractality concept was Feynman path integral approach [23],
where Feynman and Hibbs [23] reformulated the non- relativistic quantum
mechanics as a path integral over Brownian paths. Laskin [24, 25] used L\'{e}vy paths instead of the Brownian ones in the path integral and obtained
 a space fractional Schr\"{o}dinger  equation and developed the fractional quantum mechanics.

Related to fractal geometry and fractional dimensional space is the
area of fractional derivatives and integrals which have recently
been applied in many applications including particle physics [6],
fractional Hamiltonian systems [7], chaotic dynamics [4],
astrophysics [8], physics of fractals and complex media [3]
and recent studies of scaling phenomena [9,10,11]. Even though it was demonstrated in [3, 26], that the areas of fractals,
fractional dimensional space and fractional derivatives are not completely
independent, some authors have studied the Schr\"{o}dinger equation which contains fractional derivative terms (Caputo or Riesz derivatives) [27, 28, 28] in an integer space of order $n=1,2,..., $ and without taking into consideration, the fractionality dimensions of the fractals (irregularity or roughness) . For example, Dong and Xu [27], sloved the fractional Schr\"{o}dinger equation using the quantum Riesz fractional operator introduced by Laskin [24, 25].  Naber [28] showed a time Caputo fractional Schr\"{o}dinger equation. Wang and Xu [29] generalized the fractional Schr\"{o}dinger equation to construct a space- time fractional Schr\"{o}dinger equation.

Recently, Muslih and Agrawal [30] investigated the wave equation with Riesz fractional derivative and proved the connection between the Riesz derivative of order $\a$  and fractional space of order $D$ as
\begin{equation}
(-\Delta)^{\a/2}\left(\frac{1}{|\mathbf{r} -\mathbf{r'}|^{D-\a}}\right)=
\frac{ 2^{\a}\pi^{D/2}\Gamma\left(\frac{\a}{2}
\right)}{\Gamma\left(\frac{D- \a}{2}\right)}\delta^{D}(\mathbf{r}
-\mathbf{r'}),
\end{equation}
where $\delta^{D}(\mathbf{r} -\mathbf{r'})$ is the $D$  dimensional
fractional Dirac delta  function which satisfy the following identity
\begin{equation}
\int \delta^{D}(\mathbf{r} -\mathbf{r'})~d^{D}r=1
\end{equation}

Because of the fractional geometry nature of fractals, we will solve the fractional Schr\"{o}dinger equation in fractional dimensional space.  We will introduce the Fourier transform method in fractional dimensional space to solve this equation. This new method is applied to find a complete eigen solutions for a particle in a fractional $\delta$-potential well defined by $V(x) = - \gamma\delta^{\la}(x)$, ~where $\gamma>0$~ and $\delta^{\la}(x)$ is the fractional Delta function.

Our paper is organized as follows: In section two, the Fourier transform method in fractional dimensional space is presented. In section three, we introduce Fourier transform method in fractional dimensional space to solve the fractional Schr\"{o}dinger  equation. Section four deals with
the eigen solutions of a particle in fractional $\delta$-potential well. Section five contains our conclusions.

\section{Fourier transform method in fractional dimensional space}
In this section, the Fourier transform method in the spatial and momentum spaces will be established by considering the  fractional line element $d^{\la}x$  as [1, 31]
\begin{equation}
d^{\la}x = \frac{\pi^{\la/2} |x|^{\la-1}}{\Gamma(\la/2)}
dx,~ 0 < \la \le 1.
\end{equation}
The Fourier transformation is defined as [32]
\begin{equation}
g(k)= F(f(x))= \int_{-\infty}^{\infty} f(x) e^{ik x}d^{\la}x,
\end{equation}
and the inverse Fourier transformation $f(x)$ is given by
 \begin{equation}
f(x) = F^{-1}(g(k))= (\frac{1}{2\pi})^{\la} \int_{-\infty}^{\infty} g(k)e^{- ik x}d^{\la}k.
\end{equation}
This treatment leads us to the value of the generalized fractional Dirac delta function in the
$\la$ dimensional fractional space as [32]
\begin{equation}
\delta^{\la}(x-x')= (\frac{1}{2\pi})^{\la} \int_{-\infty}^{\infty}  e^{ ik (x -x')}d^{\la}k.
\end{equation}
Such, a function has the following representation [32]
\begin{equation}
\delta^{\la}(x)= \lim_{\varepsilon \rightarrow \infty}\varepsilon^{\la} e^{-\pi\varepsilon^{2} x^{2}},
\end{equation}
with the following properties\\

\begin{equation}
 \int_{-\infty}^{\infty} f(x)\delta^{\la}(x) d^{\la}x =f(0),
\end{equation}\\

\begin{equation}
\int_{-\infty}^{\infty} \delta^{\la}(x) d^{\la}x =1,
\end{equation}\\
\begin{equation}
\delta^{\la}(a x)=a^{-\la}\delta^{\la}(x).
\end{equation}\\

The Fourier convolution operator of two functions $h$ and $\varphi$ is defined by the integral
\begin{equation}\\
h \ast\varphi:= (h \ast\varphi)(x)=\int_{-\infty}^{\infty}h(x-y)\varphi(y) d^{\la}y, ~(x\in {R}).
\end{equation}\\

The Fourier transforms in spatial and momentum space can be obtained after the changing of variables from $k$ to the momentum $p= \hbar k$ as
\begin{equation}
\phi(x)= (\frac{1}{2\pi\hbar})^{\la} \int_{-\infty}^{\infty} \varphi(p) e^{- ipx/ \hbar}d^{\la}p ,
\end{equation}\\
\begin{equation}
\varphi(p)=  \int_{-\infty}^{\infty} \phi(p)e^{ ipx/ \hbar}d^{\la}x ,
\end{equation}\\
\section{Fractional Schrodinger equation in fractional space}
In this section, we will investigate the fractional Schr\"{o}dinger equation in fractional space. Laskin [24, 25], developed the fractional Feynman path integral over Levy paths using the Hamiltonian which includes the fractional kinetic term in terms of  the quantum Riesz fractional operator and is given by $(-\hbar^{2} \Delta)^{\a/2}, 1<\a\leq 2$ , where $\Delta=\frac{\p}{\p x^{2}}$ is the Laplacian in one dimension.
The time dependent Schr\"{o}dinger equation is expressed as
\begin{equation}
i\hbar \frac{\p \psi(x, t)}{\p t} = H_{\a}\psi(x, t),
\end{equation}
where $H_{\a}$ is the Laskin fractional Hamiltonian and defined as
\begin{equation}
H_{\a}= D_{\a}(-\hbar^{2} \Delta)^{\a/2} + V(x, t).
\end{equation}
Here, $D_{\a}$ has physical dimension $[D_{a}]= erg^{1-\a}\times cm^{\a}\times s^{-\a}$~$(D_\a =\frac{1}{2m}$ ~for $\a=2$, $m$ is the physical mass of the particle).

For  Hamiltonians do not depend explicitly on time and taking into account that $\psi(x, t)= \phi(x)e^{- i E/\hbar}$, then Eq. (14) can be put in the form.
\begin{equation}
D_{a}(-\hbar^{2} \Delta)^{\a/2} \phi(x) + V(x)  \phi(x) = E \phi(x),
\end{equation}
As was specified in the introduction, the solutions of time dependent and independent fractional Schr\"{o}dinger equations in the integer dimensional space, are studied by many some [27-29]. Our aim is to use the method introduced in the previous section and to investigate the eigen solutions of fractional Schr\"{o}dinger in fractional dimensional space. As an example we will solve a fractal source with a potential in the form of fractional Dirac delta distribution function.
To accomplish this goal, we multiply Eq. (16) from left by by $e^{-ipx/\hbar}$ and taking the Fourier transform of the resultant value, and then integrating over $x$
from $-\infty$ to $\infty$ we have,
\bn
&&\int_{-\infty}^{\infty} \left(D_{\a} |p'|^{\a}\varphi(p')d^{\la}p' \left\{\int e^{ix/h(p'-p)}d^{\la}x\right\}\right) + \nonumber\\
&&\int_{-\infty}^{\infty} e^{-ix/h(p-p')} V(x)d^{\la}x \varphi(p')d^{\la}p'=
E \int_{-\infty}^{\infty} e^{-ix/h(p-p')} d^{\bt}x \varphi(p')d^{\la}p'.
\en
Using the definition of fractional Dirac delta function (6), we obtain
\begin{equation}
(2\pi\hbar)^{\la}(D_{\a} |p|^{\a}\varphi(p))  + \int_{-\infty}^{\infty} e^{-ix/h(p-p')} V(x)d^{\la}x \varphi(p')d^{\la}p'= (2\pi\hbar)^{\la} E \varphi(p).
\end{equation}
\section{Fractional $\delta$-potential}
In this section, we consider a particle in fractional $\delta$-potential well defined by $V(x) =- \gamma\delta^{\la}(x)$, ~$\gamma>0$~ and $0<\la \leq 1$, where $\delta^{\la}(x)$ is the fractional Dirac delta function defined in Eq. (7) . The time independent fractional Schr\"{o}dinger equation for this particle is given by
\begin{equation}
D_{a}(-\hbar^{2} \Delta)^{\a/2} \phi(x)  - \gamma\delta^{\la}(x) \phi(x) = E \phi(x).
\end{equation}
We consider here $E<0$. Making use of Eq. (18), we obtain
\begin{equation}
(2\pi\hbar)^{\la}(D_{\a} |p|^{\a}\varphi(p))  + \int_{-\infty}^{\infty} e^{-ix/h(p-p')} ( - \gamma\delta^{\la}(x))d^{\la}x \varphi(p')d^{\la}p'= (2\pi\hbar)^{\la} E \varphi(p).
\end{equation}
Using the property (8), we obtain
\begin{equation}
(D_{\a} |p|^{\a}\varphi(p))  -\frac{\gamma}{(2\pi\hbar)^{\la}} \int_{-\infty}^{\infty} \varphi(p)d^{\la}p= E \varphi(p).
\end{equation}
Let
\begin{equation}
\int_{-\infty}^{\infty} \varphi(p)d^{\la}p=C,
\end{equation}
where $C$ is a constant. Substituting Eq. (22) in Eq. (21), we have
\begin{equation}
\varphi(p)= \frac{ - \gamma}{(2\pi\hbar)^{\la}}\frac{C}{D_{\a} |p|^{\a}-E}.
\end{equation}
Again, substitution of the solution (23) in Eq. (22), we obtain
\begin{equation}
\int_{-\infty}^{\infty} \frac{d^{\la}p}{D_{\a} |p|^{\a}-E}= \frac{(2\pi\hbar)^{\la}}{\gamma}.
\end{equation}
Making use of the fractional line element defined in Eq. (3), then Eq. (24) can be put in the form
 \begin{equation}
\frac{2\pi^{\la}/2}{\Gamma(\la/2)}\int_{0}^{\infty} \frac{|p|^{\la -1}dp}{D_{\a} |p|^{\a}-E}= \frac{(2\pi\hbar)^{\la}}{\gamma}.
\end{equation}
Using the identity [33]
\begin{equation}
\frac{z^{\la}}{1 + a z^{\a}} = a^{\la/\a}H^{1,1}_{1,1}\left[az^{\a}|^{(\la/\a, 1)}_{(\la/\a,
1)}\right].
\end{equation}
where $ H^{m,n}_{p,q}(z)$ is the H Fox-function (for more details, see the Appendix), we obtain
\begin{equation}
\frac{|p|^{\la -1}}{D_{\a} |p|^{\a}-E} =-\frac{1}{E}(\frac{D_{\a}}{-E})^{-(\frac{\la -1}{\a})}H^{1,1}_{1,1}\left[(\frac{D_{\a}}{-E})|p|^{\a}|^{(\la-1)/\a, 1)}_{(\la-1)/\a,
1)}\right].
\end{equation}
This allows us to calculate the integral
\begin{equation}
\int_{0}^{\infty} \frac{|p|^{\la -1}dp}{D_{\a} |p|^{\a}-E}= -\frac{1}{\a E}\left(-\frac{E}{D_{\a}}\right)^{(\frac{\la-2}{\a})}\int_{0}^{\infty} H^{1,1}_{1,1}\left[(\frac{D_{\a}}{-E})^{1/\a}|p|^{\a}|^{(\la-1)/\a, 1/\a)}_{((\la-1)/\a, 1/\a)}\right]~dp,
\end{equation}
Using the identity (41),  we obtain the the eigen values $E_{\a, \la}$ for the particle in fractional $\delta$-potential well as
\begin{equation}
E_{\a, \la}=- \left(\frac{\gamma \Gamma(\la/\a)\Gamma(1- \la/\a)}{2^{\la-1}\hbar^{\la}\pi^{\la}\Gamma(\la/2)\a (D_{a})^{\la/\a}}\right)^{\frac{\a-\la}{\a}}.
\end{equation}
According to the identity (41), the integral (28) exists, when $ 0<\la< \a$. For one dimensional fractal systems $0<\la\leq 1$ we obtain $1<\a$. As special cases, for $\la=1, ~\a=2$, the value of $E_{\a, \la}$ reduces to the same energy eigen value as given in standard mechanics [34, 35].

Making use of Eqs. (28) and (12), we obtain the solution $\phi(x)$ as

 \bn
\phi(x)&&= \frac{\gamma C}{(2\pi\hbar)^{2\la}}\int_{-\infty}^{\infty} \frac{e^{ipx/\hbar}}{D_{\a} |p|^{\a}-E}d^{\la}p\nonumber\\
&&= \frac{\gamma C}{(2\pi\hbar)^{2\la}}\frac{2\pi^{\la}/2}{\Gamma(\la/2)}\int_{0}^{\infty}\frac{e^{ipx/\hbar}|p|^{\la-1}}{D_{\a} |p|^{\a}-E}dp\nonumber\\
&&=- \frac{\gamma C}{(2\pi\hbar)^{2\la}}\frac{2\pi^{\la}/2}{\Gamma(\la/2)E}\int_{0}^{\infty}\frac{e^{ipx/\hbar}|p|^{\la-1}}{1- D_{\a}/E |p|^{\a}} dp.
\en
Again, using the identity formula (26), we obtain
\begin{equation}
\phi(x)= - \frac{\gamma C}{(2\pi\hbar)^{2\la}}\frac{2\pi^{\la}/2}{\Gamma(\la/2)E}\int_{0}^{\infty}e^{ipx/\hbar}(\frac{D_{\a}}{-E})^{-(\frac{\la -1}{\a})}H^{1,1}_{1,1}\left[(\frac{D_{\a}}{-E})|p|^{\a}|^{(\la-1)/\a, 1)}_{((\la-1)/\a,
1)}\right]dp.
\end{equation}
The integral (31) can be evaluated using the Fourier cosine transform of the $ H$ function [36] as follows:
\begin{equation}
\phi(x)= C_{\a}^{\la} F_{\a}^{\la}(x),
\end{equation}
where $C_{\a}^{\la}$ and  $F_{\a}^{\la}(x)$ are given respectively as
\begin{equation}
 C_{\a}^{\la}= \frac{\gamma C}{(2\pi\hbar)^{2\la}}\frac{2\pi^{\la}/2}{\Gamma(\la/2)}\frac{1}{(D_\a)^{\frac{\la-1}{\a}}(-E)^{\frac{\a + 1 -\la }{\a}}},
\end{equation}
\begin{equation}
F_{\a}^{\la}(x)= \int_{0}^{\infty}\cos(px/\hbar)H^{1,1}_{1,1}\left[(\frac{D_{\a}}{-E})|p|^{\a}|^{(\la-1)/\a, 1)}_{((\la-1)/\a,
1)}\right]dp.
\end{equation}

With the help of (44), $F_{\a}^{\la}(x)$ can be evaluated as
\begin{equation}
F_{\a}^{\la}(x)= \frac{\pi \hbar}{|x|}H^{2,1}_{2,3}\left[(\frac{|x|}{\hbar})^{\a}(\frac{-E}{D_{\a}})|^{(\frac{\a-\la+1}{\a}, 1), (1, \a/2)}_{(1, \a/2), (\frac{\a-\la+1}{\a}, 1), (1, \a/2)}\right].
\end{equation}

Using  (45), Eq. (35), reduces to
 \begin{equation}
F_{\a}^{\la}(x)= \frac{\pi \hbar}{\a}H^{2,1}_{2,3}\left[|x|(\frac{D_{\a}{\hbar}^{\a}}{-E})^{-1/\a}|^{(\frac{\a-\la}{\a}, 1/\a), (1/2, 1/2)}_{(0, 1), (\frac{\a-\la}{\a}, 1/\a), (1/2, 1/2)}\right].
\end{equation}

Furthermore,  using the formulas (46) and (47), we obtain
 \begin{equation}
F_{\a}^{\la}(x)= \frac{\pi \hbar}{\a}H^{1,0}_{0,1}\left[|x|(\frac{D_{\a}{\hbar}^{\a}}{-E})^{-1/\a}|(0, 1)\right].
\end{equation}
Finally we obtain the wave function $\phi(x)$ for a particle in fractional $\delta$-potential well as
\begin{equation}
\phi(x)= C_{\a}^{\la}\frac{\pi \hbar}{\a}H^{1,0}_{0,1}\left[|x|(\frac{D_{\a}{\hbar}^{\a}}{-E})^{-1/\a}|(0, 1)\right].
\end{equation}
For the special case $\a=2$ and $\la=1$, the wave function in Eq. (38), reduces to that in the standard quantum mechanics [34, 35].
\section{Conclusions}
In this paper we developed the Fourier transform in fractional space. This allows us solve fractional Schr\"{o}dinger in fractional dimensional space $\la$. As an example we obtained  a complete eigen solutions for a particle in a fractional $\delta$-potential in the form $V(x) =- \gamma\delta^{\la}(x)$. ~where $\gamma>0$~ and $\delta^{\la}(x)$ is the fractional Dirac delta function.
 Using the properties of Fox H-functions we demonstrated that the eigen solutions are exist if $0< \la<\a$.

\section*{Appendix}
In this appendix, we will review the Fox's H-function, and its properties, which have been used in our calculations. The  Fox's H-function is defined by the contour integral, [33]
\begin{equation}
H^{m,n}_{p,q}(z)=H^{m,n}_{p,q}\left[z|^{(a_p,A_p)}_{(b_q,
B_q)}\right]= \frac{1}{2\pi i}\int_{L} h(s)z^{s}~ds,
\end{equation}
where $h(s)$ is given by
\begin{equation}
h(s)= \frac{\displaystyle\prod_{j=1}^{m}\Gamma(b_j + B_j
s)\displaystyle\prod_{j=1}^{n}\Gamma(1-a_j +A_j s)}
{\displaystyle\prod_{j=m+1}^{q}\Gamma(1-b_j +B_j
s)\displaystyle\prod_{j=n+1}^{p}\Gamma(a_j + A_j s)}.
\end{equation}
Where $ m, n, p, q$ are integers satisfying $m^2 + n^2 \neq0$,~ $0 \leq n \leq p$,~ $0  \leq m \leq q$

 and empty products are interpreted as unity. The parameters $a_j(j=1,...,p)$ and $b_j(1,...,q)$ are complex numbers and $A_j(j=1,..,p)$ and

$B_j(j=1,...,q)$ are positive numbers satisfying $P_a \cap P_b =\emptyset$, where $P_a= \{s=(b_j+k)/B_j, ~j=1,2,...,m;k=0,1,2,...\}$~ and
$P_b= \{s=(A_j-1-k)/A_j, ~j=1,2,...,m;k=0,1,2,...\}$ .
The integration contour runs from $s=c-i\infty$ to $s=c+i\infty$ such that $P_a$ lies to right of $L$ and $P_b$ to the left of $L$.
The Mellin transform of a single H-function is given as

\bn \label{Mit4}
&&\int_{0}^{\infty}z^{s-1}H^{m,n}_{p,q}\left[az|^{(a_p,A_p)}_{(b_q,
B_q)}\right]dz\nonumber\\
&&= a^{-s}\frac{\displaystyle\prod_{j=1}^{m}\Gamma(b_j + B_j
s)\displaystyle\prod_{j=1}^{n}\Gamma(1-a_j +A_j s)}
{\displaystyle\prod_{j=m+1}^{q}\Gamma(1-b_j +B_j
s)\displaystyle\prod_{j=n+1}^{p}\Gamma(a_j + A_j s)}, \en where $-
min_{1 \leq j\leq m}~ R(\frac{bj}{B_j})< R(s)<\frac{1}{A_j}-
max_{1 \leq j\leq  n} R(\frac{aj}{A_j}), |arg~a|<\frac{1}{2}\pi\la$,\\
$\la =\displaystyle\sum_{j=1}^{m}A_{j}-
\displaystyle\sum_{j=n+1}^{p}A_{j}
\displaystyle\sum_{j=1}^{m}B_{j} -
\displaystyle\sum_{j=m+1}^{q}B_{j}>0$.

Now we would like to derive the cosine transform of H-function.
Using the integral [33]
\bn
&&\int_{0}^{\infty}z^{s +1/2}J_{\nu}(kz)H^{m,n}_{p,q}\left[az^{\mu}|^{(a_p,A_p)}_{(b_q,
B_q)}\right]dz= 2^{s +1/2}k^{-s-3/2}\times\nonumber\\&&H^{m,n+1}_{p+2,q}\left[(2/k)^{\mu}z|^{(1/4-s/2-\nu/2, \mu/2),(a_p,A_p),(1/4-s/2+\nu/2, \mu/2)}_{(b_q,
B_q)}\right],
\en
and the values of cosine function
\begin{equation}
J_{-1/2}(kz)=\sqrt{\frac{2}{\pi kz}}\cos{kz},
\end{equation}
we obtain the cosine transform of H-function as
\bn
&&\int_{0}^{\infty}z^{s -1}\cos(kz)H^{m,n}_{p,q}\left[az^{\mu}|^{(a_p,A_p)}_{(b_q,
B_q)}\right]dz=\nonumber\\&& \frac{\pi}{k s}H^{m+1,n}_{q+1, p+2}\left[k^{\mu}/a|^{(1-bq, Bq), (\frac{1+s}{2}, \mu/2)}_{(s, \mu), (1-aq, Aq), (\frac{1+s}{2}, \mu/2)}\right].
\en
Some properties of H-function
\begin{equation}
H^{m,n}_{p,q}\left[z^{\mu}|^{(a_p,A_p)}_{(b_q,
B_q)}\right]= \frac{1}{\mu}H^{m,n}_{p,q}\left[z|^{(a_p,\frac{A_p}{\mu})}_{(b_q,
\frac{B_q)}{\mu}}\right], ~~~\mu>0.
\end{equation}
\bn
&&H^{m,n}_{p,q}\left[z|^{(a_1, A_1), (a_2, A_2),...,(a_p,A_p)}_{(b_1, B_1), (b_2, B_2),..., (b_{q-1}, B_{q-1}, (a_1, A_1)}\right]=\nonumber\\&&
H^{m,n-1}_{p-1,q-1}\left[z|^{(a_2, A_2),...,(a_p,A_p)}_{(b_1, B_1), (b_2, B_2),..., (b_{q-1}, B_{q-1})}\right].
\en
\bn
&&H^{m,n}_{p,q}\left[z|^{(a_1, A_1), (a_2, A_2),...,(b_1, B_1)}_{(b_1, B_1), (b_2, B_2),..., (b_{q}, B_{q})}\right]=\nonumber\\&&
H^{m-1,n}_{p-1,q-1}\left[z|^{(a_1, A_1),...,(a_{p-1},A_{p-1})}_{(b_2, B_2),..., (b_{q}, B_{q})}\right].
\en

\end{document}